\documentclass[letterpaper
,aps,pra
,twocolumn
,showkeys
]{revtex4-2}

\usepackage{graphicx}% Include figure files
\usepackage{dcolumn}% Align table columns on decimal point
\usepackage{bm}% bold math
\usepackage{latexsym}
\usepackage{ifsym}
\usepackage{stix}% Geometrical shapes  DOES NOT WORK with {fourier}
\usepackage[utf8]{inputenc}
\usepackage{amsmath}
\usepackage{longtable}
\usepackage{ltxtable}
\usepackage{booktabs}

\begin{document}

\bibliographystyle{apsrev}

\title{On the principal series of potassium}

\author{Constantine E. Theodosiou}
\affiliation{Department of Physics and Astronomy, Manhattan College, Riverdale, New York 10471}
\email{E-mail: constant.theodosiou@manhattan.edu}

\date{\today}        

\begin{abstract}
We revisit the photoabsorption from the ground state of K through the first ionization limit, and provide recommended values for the optical oscillator strengths.  The anomaly in the oscillator strength ratio within each doublet is also revisited and compared with both the experimental data available and the relativistic calculations available.   
\end{abstract}

\pacs{...}

\maketitle

\section{Introduction} 
The ground-state photoabsorption cross sections of alkali metal atoms have been studied for a long time now, but the measurements are hampered by several factors, both for discrete transitions as well as for ones to the continuum. High Rydberg states are greatly affected by Stark effect from stray fields and neighboring atoms, whereas the photoionization cross section near thresholds displays a deep minimum (Cooper minimum \cite{Cooper1962, Fano1968}) for all alkalis but lithium, and the presence of even minute amounts of molecular species in the target area masks the presence of the minima. There has been a lot of work in calculating the photoionization cross section using a vast array of methods with varying degrees of success.  The calculation of oscillator strengths to states $np$ with $n$ larger than 10, has not been addressed by self-consistent \textit{ab initio} multiconfiguration and/or relativistic calculations, and has been dominated by semiempirical or model potential calculations that use the experimental energy as input parameters to obtain appropriate central fields and wavefunctions.  The atomic cores of all alkali atoms exhibit prominent polarizability effects and the explicit inclusion of these effects has been a prerequisite to obtaining reasonable results.  Even the most sophisticated calculations utilize cutoff radii for the expression of the polarizability terms in the wavefunction and matrix elements calculated.  Dalgarno and coworkers \cite{Weisheit1972} have initiated a series of fundamental theoretical studies of the polarization effects on the atomic states and the transition rates between them.  That work has evolved greatly over the recent five decades.

As just mentioned, the availability of experimental optical oscillator strength values for medium to high Rydberg states is extremely limited and most data go back several decades.  New experimental techniques, using lasers and multiphoton processes have opened new opportunities for measuring those states and some recent work has appeared in the literature.  Still, there are not enough -- in some cases any -- comparisons with dependable theoretical results to establish the veracity of the data extracted form such measurements.  Thus, it becomes necessary to develop dependable calculation methods that provide oscillator values to better than  5\% accuracy, so that benchmark values can be established.  This is the intent of the present work.  

This paper addresses the peculiar case of the potassium principal series ($4s$$\rightarrow$$np$).  The results of early measurements have been tabulated by the U.S. National Bureau of Standards (NBS/NIST) \cite{Wiese1969}.   The early work of Mazing and Serapinas \cite{Mazing1969} provides the first available data for $n$ in the 18 to 30 range and claim a 20\% accuracy.  Measurements by Shabanova and Khlyustalov \cite{Shabanova1984} have also given a useful trend of the measured values at lower $n$, with a 20\% accuracy. The work of Huang and Wang \cite{Huang1981} produced relative oscillator strengths for $n=7-50$, with claimed 10\% accuracy.  They also claimed for $n=9$ absolute oscillator strengths for $f_{3/2}$ and $f_{1/2}$ and a doublet value of absolute accuracy of 6\%.  Newer measurements include a couple of efforts by Connerade and coworkers \cite{Nawaz1992}, but their (relative) measurements have been plagued by magnetic field disturbances and, as we will see, do not seem to give us any guidance on the experimental trend of these quantities. The former three measurements, combined with some accurate measurement of the photoionization cross section at the ionization threshold, could give a definitive answer to the correct values of the oscillator strengths.  Unfortunately, even at threshold, the experimental photoionization results deviate form each other, or are not absolute in value.  Several, presumed accurate, theoretical calculations have been made in recent years, which also disagree on the exact position of the (Cooper) minimum and the value at threshold, but are closer together than the experimental ones.  A relatively recent relative measurement of the photoionization cross section near the Cooper minimum by Sandner et al.\ \cite{Sandner1981} provides an important benchmark point about the location of the minimum \underline{and} its width.  So, even though the measurement was relative, a theory that reproduces these two characteristics ought to be an accurate one and its threshold photoionization value can be taken as a good reference point, so that the  optical oscillator strengths can be also calibrated (if they are relative in value).  A more subtle effect of the alkali metals is that their doublet ratio of oscilllator strengths $\rho=f(4s-np_{3/2})/f(4s-np_{1/2})$ deviates greatly from the the expected statistical ratio of 2, due to the spin-orbit interaction.  Thus,  correctly predicting or measuring this ratio provides an added criterion on the accuracy of the method, experimental or theoretical.

The work of this investigator, its method being briefly outlined below, has been successful in treating Rydberg state lifetimes \cite{CET1984} and, more recently \cite{CET2022}, in accurately and self-consistently predicting/confirming the location, depth, and width of the Cooper minimum of the K($4s$) photoionization.  Therefore, it can serve as such a reference point and enable us to provide recommended values for the entire principal series oscillator strengths. The present treatment is compared below with all available (published) experimental data and with several theoretical treatments, some of them considered very accurate -- but limited to lower principal quantum numbers $n$ -- and resolves the Cooper minimum width discrepancy through a careful treatment of the spin-orbit interaction and core-polarization effects. 

\section{Method of Calculations}
The wave functions are obtained by solving the Schr\"odinger equation
\begin{equation}
\left[ \frac{d^2}{dr^2}-V(r)-\frac{l(l+1)}{r^2}+E_{nl}\right]
P_{nl}(r)=0.
\end{equation}
The central potential
\begin{equation}
V(r)=V_{{\rm HKS}}(r)+V_{{\rm pol}}(r)+V_{{\rm so}}(r)=V_{{\rm m}}(r)+V_{%
{\rm so}}(r)
\end{equation}
consists of three terms: $V_{\rm HKS}(r)$, a Hartree-Kohn-Sham-type \cite{Desclaux1969} self-consistent field term,
\begin{equation}
V_\text{pol}(r)=
-\frac 12\frac{\alpha _d}{r^4}\left\{ 1-{\rm exp}
\left[-(r/r_c)^6\right] \right\}                             
\end{equation}
a core-polarization term, and
\begin{equation}
V_{{\rm so}}(r)=-\frac 12\alpha ^2\left\{ 1+\frac{\alpha ^2}
4[E-V(r)]\right\} ^{-2}\frac 1r\frac{dV_{\rm m}(r)}{dr}\vec L\cdot \vec S
\end{equation}
a spin-orbit interaction, Pauli approximation term. Here $\alpha_d$ is the dipole 
polarizability of the core \cite{Johnson1981}, $r_c$ is a cut-off distance and $\alpha$ is the fine-structure constant.

The necessary radial matrix elements were calculated using the modified dipole operator expression similar to the one used by Norcross \cite{Norcross1973},
\begin{subequations}
\begin{align}
R(nl,n'l'j')=\left< n'l'j'\left| r_\text{mod}\right| nlj \right >,%\\
\end{align}
\begin{align}
r_\text{mod}= r \left[ 1-\frac{\alpha _d}{r^3}\left( 1-\frac{1}{2} \left[ e^{\left\{ -(r/r_{cl})^3\right \}}
 +e^{\left \{-(r/r_{cl^{\prime}})^3 \right \}} \right ] \right) \right ]. \ \ \ \ 
\end{align}
\end{subequations}

The cutoff distances $r_{cl}$ are taken to be equal to the values used in the polarization potential $V_\text{pol}$ needed to reproduce the lowest experimental energy for each symmetry; they are different for each value of $l$.  They are the only adjustable parameters in this approach.

The photoabsorption cross section is obtained using \cite{Sobelman1992}
\begin{equation}
\sigma(nlj\rightarrow n'l'j')=\frac43 \pi^2a_0^2\alpha
(\epsilon_{n'l'j'}-\epsilon_{nlj})\frac{1}{2j+1} S(nlj,n'l'j')
\end{equation}
where energy is given in Rydbergs, and the absorption oscillator strength $f$ is 
\begin{equation}
f(nlj\rightarrow n'l'j')=\frac23(\epsilon_{n'l'j'}-\epsilon_{nlj})
\frac{1}{2j+1}S(nlj,n'l'j')
\end{equation}
The line strength $S(nlj,n'l'j')$ %is the same as the reduced dipole matrix element and 
in the case of alkali metal atoms is given by
\begin{equation}
\begin{aligned}
S(nlj,n'l'j')=&\text{max}(l,l')(2j+1)(2j'+1)\\
&\times
\begin{Bmatrix}  
l &{\ } l' &{\ } 1 \\
j' &{\ } j &{\ } \frac12 \\
\end{Bmatrix}%}
^2 R(nlj,n'l'j')^2.
\end{aligned}
\end{equation}

When the final state is in the continuum, $\left|n'l'j'\right>$ is replaced by $\left|\epsilon l'j'\right>$ and the total photoionization cross section of an initial state $\left|nlj\right>$ to the continuum comprises from
the sum of the partial cross sections for a photon energy $E=\epsilon-\epsilon_{nlj}$:
\begin{equation}
\begin{aligned}
\sigma_{nlj}(E)=&\frac43 \pi^2a_0^2\alpha (\epsilon-\epsilon_{nlj}) \sum_{l'j'} \text{max}(l,l') (2j'+1)\\
&\times
\begin{Bmatrix}  
l &{\ } l' &{\ } 1 \\
j' &{\ } j &{\ } \frac12 \\
\end{Bmatrix}
^2
R(nlj,\epsilon l'j')^2.
\end{aligned}
\end{equation}

The continuum photoionization cross section joins smoothly at the energy threshold with the oscillator strength distribution for discrete states using the formula
\begin{equation}
\sigma(nlj\rightarrow{}n'l'j')=\frac{2\pi^2\alpha\hbar^2}{m} \frac{df}{dE}\\
\end{equation}
where
\begin{equation}
\frac{df}{dE}\equiv \frac{df_{n'l'j'}}{d\epsilon_{n'l'j'}} =\frac{1}{2}(n_{j'}^*)^3
f(nlj\rightarrow n'l'j').
\end{equation}
\noindent
This smooth joining is very useful in normalizing relative measurements on either side of the threshold.

\section{Results and Discussion}

\begin{figure}[h]\centering\small\label{K-4s_cs_m}
\includegraphics[width=8.6cm]{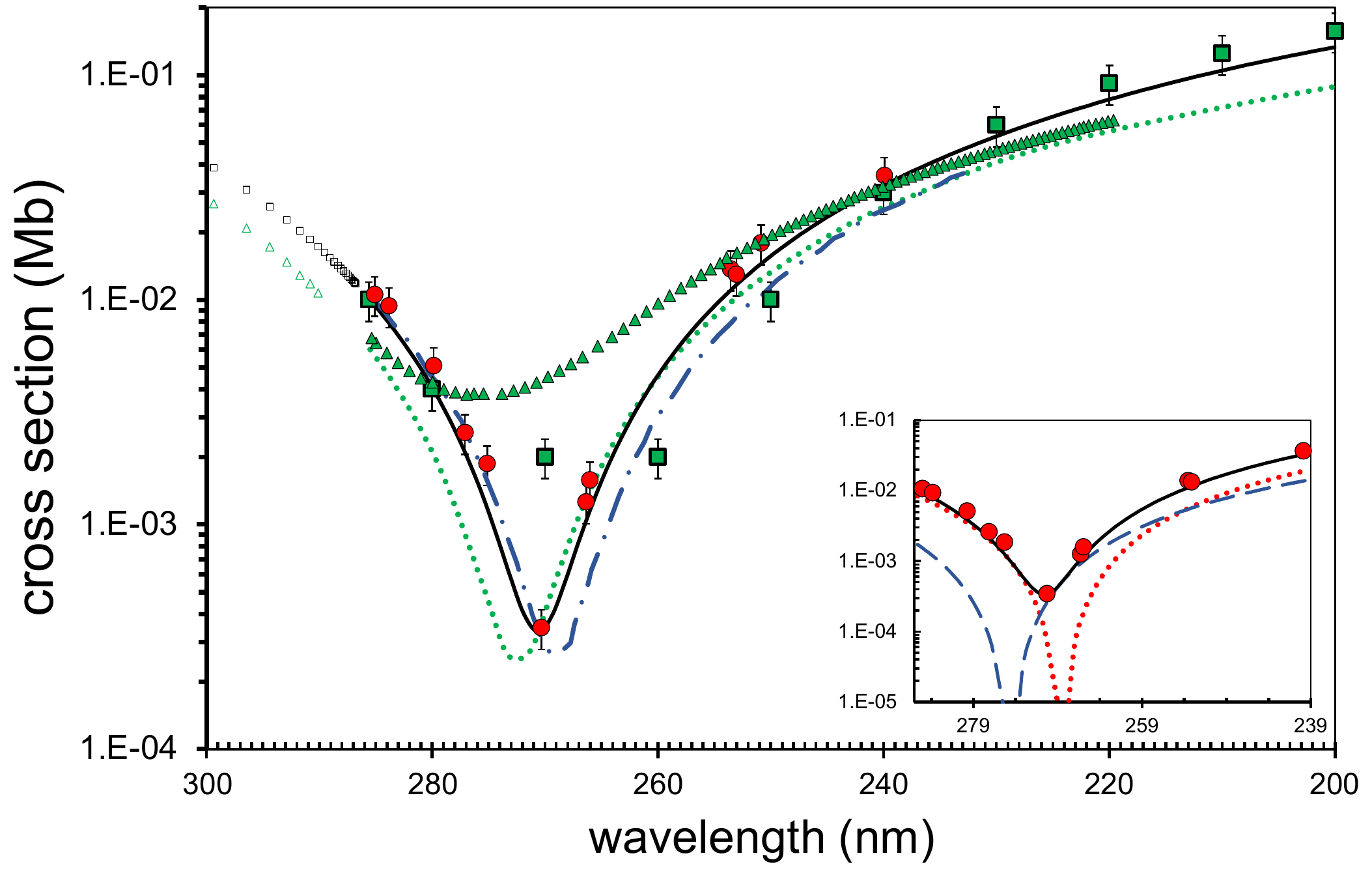}
\caption{\label{fig: K-4s_cs_m} (Color online) K($4s$) photoabsorption cross section around the first threshold.
Experimental: $\mdblkcircle$, Sandner et al.\cite{Sandner1981}, $\mdwhtsquare$, Hudson and Carter\cite{Hudson1965}, $\bigtriangleup$, Marr and Creek\cite{Marr1968}.
Theoretical: dots (green), Weisheit\cite{Weisheit1972}; 
dash-dot (blue), Zatsarinny and Tayal\cite{Zatsarinny2010}; solid line, present results.  The respective data below the threshold represent the optical oscillator strengths. \textit{Insert:} Partial-wave contributions to the K($4s$) photoionization cross section.  $\mdblkcircle$, Sandner et al.\ \cite{Sandner1981}; dots (red), $4s$$\rightarrow$$\epsilon p_{1/2}$ cross section; dashes (blue),  $4s$$\rightarrow$$\epsilon p_{3/2}$ cross section; solid line, total cross section. [From Ref.\ \cite{CET2022}]}
\end{figure}

Fig.\ \ref{fig: K-4s_cs_m} displays the results of our calculations for the K($4s$) photoabsorption, along with those by Weisheit \cite{Weisheit1972}, used a standard reference in the literature, and the more recent work of Zatsarinny and Tayal \cite{Zatsarinny2010}, and compares them with the available experimental results near threshold \cite{Hudson1965, Marr1968, Sandner1981}.  From our excellent agreement with Sandner et al.\ \cite{Sandner1981} we can safely establish the threshold ionization cross section at 0.010(1) Mb, and the ionization limit value of $df/dE$ at $8.49\times 10^{-5}$ [eV]$^{-1}$.

%$$$$$$$$$$$ fig.1 $$$$$$$$$$$$$$$$$$$$$$$$$$$$$

Our calculations extend from photoabsorption to high Rydberg states below the first ionization threshold, to continuum states just above the well-known Cooper minimum above threshold.
All three calculations reproduce the general shape of the minimum but vary in the prediction of its location. Our results reproduce best the data by Sandner et al.\ \cite{Sandner1981}.
The older data by Hudson and Carter \cite{Hudson1965} reproduce the ``wings'' well, but are not accurate around the the bottom of the minimum. The data Marr and Creek \cite{Marr1968} display a shallow minimum and grow slower with increasing energy.

%$$$$$$$$$$ fig.2 $$$$$$$$$$$$$$$$$$$$$$$$$$$$$
\begin{figure}[h]\centering\small\label{Fig2}
\includegraphics[width=7.0cm]{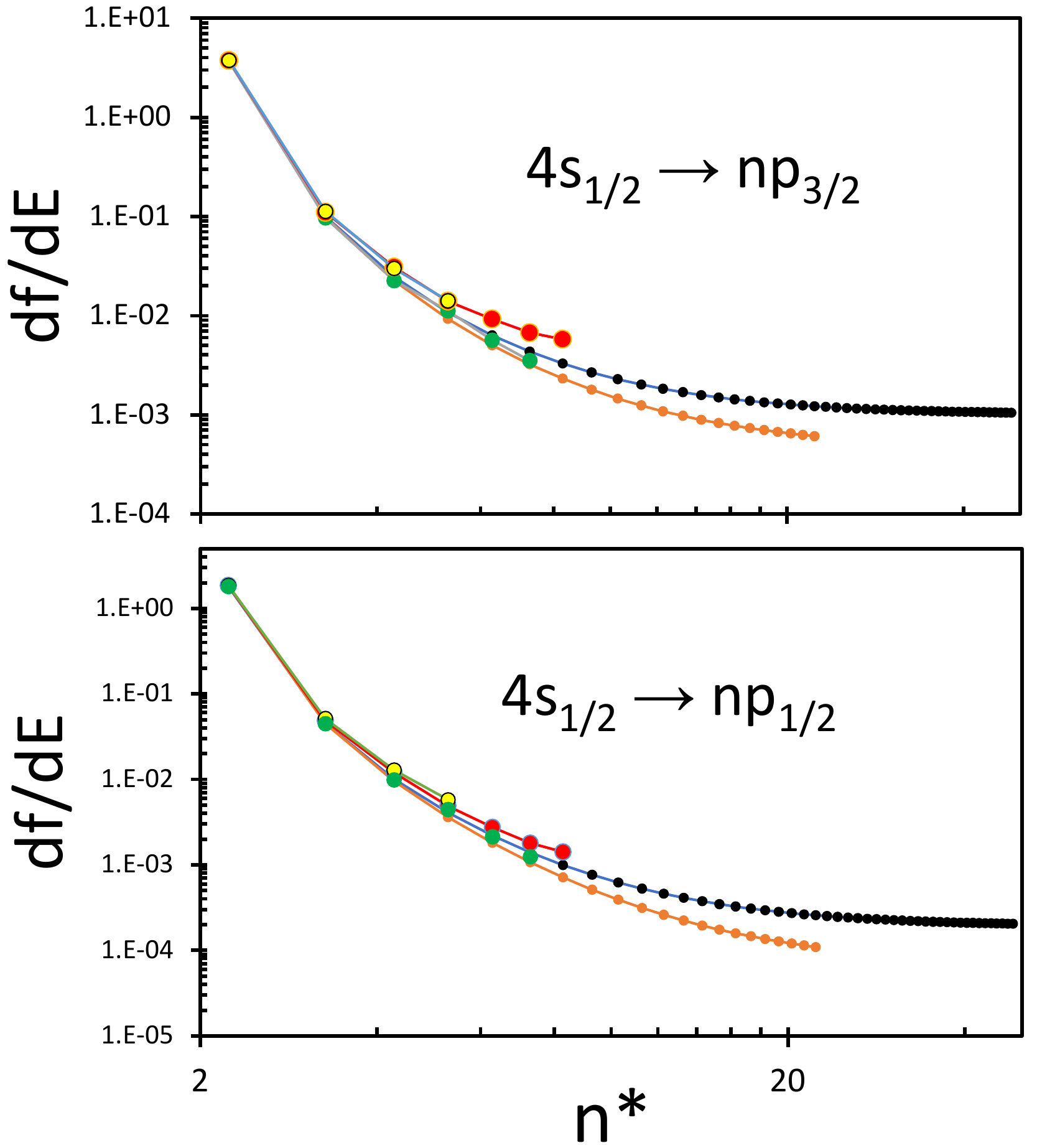}
\caption{\label{fig: K-4s-np1_3_df_dE}   (Color online) K $4s-np$ absorption oscillator strengths. 
\textit{Top}: $4s_{1/2}-np_{3/2}$;  \textit{Bottom}: $4s_{1/2}-np_{1/2}$. 
\textit{Experimental:}  $\mdblkcircle$ (green), Shabanova and Khlyustalov \cite{Shabanova1984}. 
\textit{Theoretical:} $\mdblkcircle$ (red), using the matrix elements of Safronova et al.\ \cite{Safronova2013}; $\mdblkcircle$ (yellow), Nandy et al.\ (2012);  $\mdblkcircle$ (orange), Migdalek and Kim\cite{Migdalek1998}; solid black line with bullets, present.}  
\end{figure} 

%$$$$$$$$$$ fig.3 $$$$$$$$$$$$$$$$$$$$$$$$$$$$$
\begin{figure}[h]\centering\small\label{Fig3}
\includegraphics[width=8.6cm]{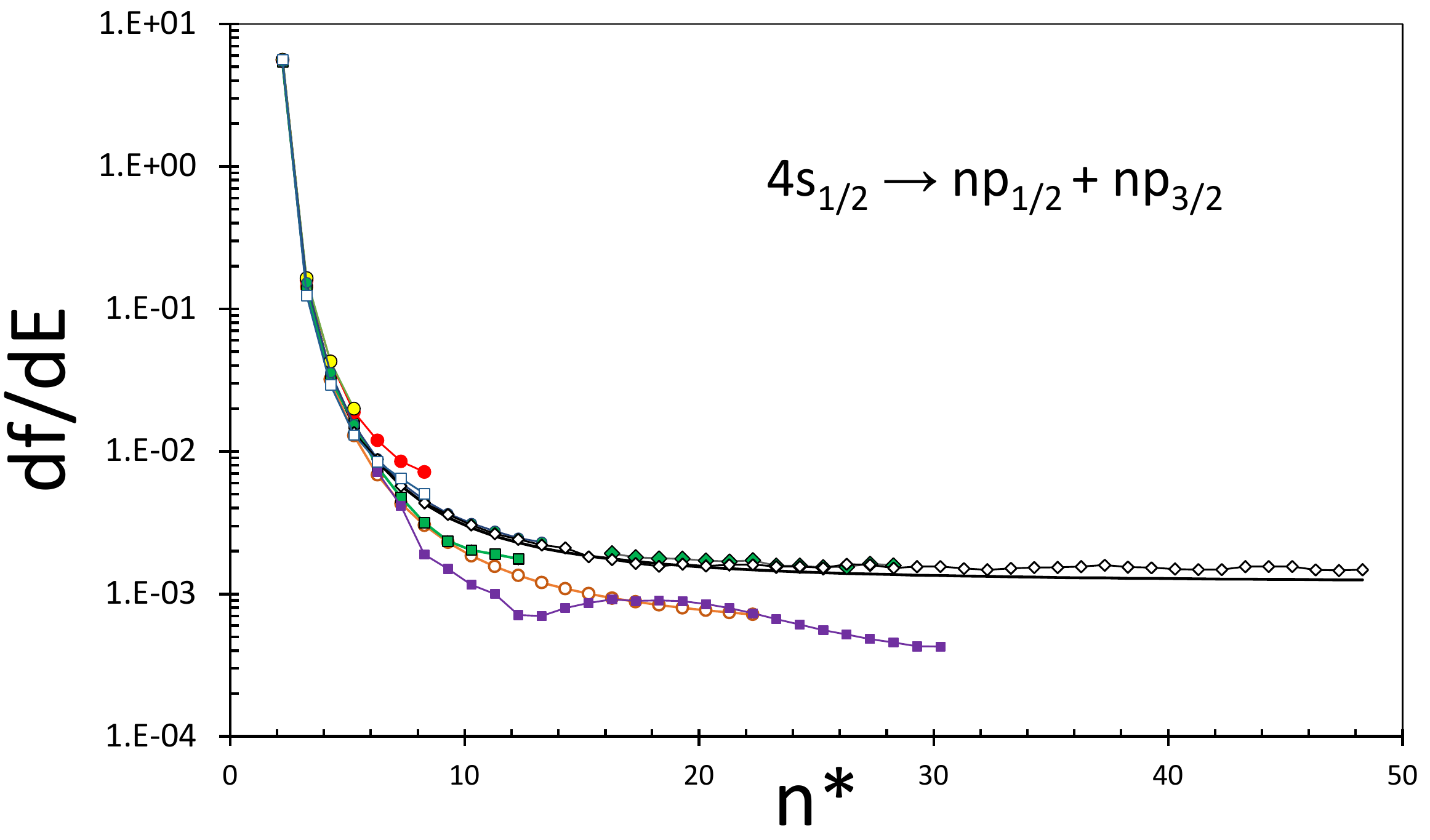}
\caption{\label{fig: K-4s-np_df_dE}   (Color online)  
K $4s_{1/2}-np_{3/2}+np_{1/2}$ absorption oscillator strengths. 
\textit{Experimental:}  $\mdwhtsquare$, Filippov \cite{Filippov1933}; $\mdblksquare$ (green), Shabanova and Khlyustalov \cite{Shabanova1984}; $\mdblkdiamond$ (green), Mazing and Serapinas \cite{Mazing1969}; $\mdwhtdiamond$, Huang and Wang \cite{Huang1981}.   
\textit{Theoretical:} $\mdblkcircle$ (red), using the matrix elements of  Safronova et al.\  \cite{Safronova2013}; $\mdblkcircle$ (yellow), Nandy et al.\ (2012);  $\mdblkcircle$ (orange), Migdalek and Kim\cite{Migdalek1998}; $\mdblkcircle$ (green), Weisheit \cite{Weisheit1972}; solid black line with bullets, present.}  
\end{figure}

Figs.\ \ref{fig: K-4s-np1_3_df_dE} and \ref{fig: K-4s-np_df_dE} present the collective results, in the form of $df/dE$, that clearly display the asymptotic behavior toward the ionization threshold.  The agreement with the values of Mazing and Serapinas \cite{Mazing1969} is excellent and particularly gratifying, since the measurements were made at relatively hign $n$ values.  The data of Shabanova et al.\ \cite{Shabanova1984} agree well with our predicted values, though they are lower at higher $n$.  The relative data of Nawaz et al.\ \cite{Nawaz1992} have a different behavior and do not seem to converge to a uniform value towards the threshold; those measurements were affected, according to their authors, by magnetic field effects and were renormalized in an effort to remove those effects.  The data of Huang and Wang \cite{Huang1981} cover a wide range of $n$.  They were relative and were  converted to absolute by using an absolute measurement of the 9$p$ doublet components and a doublet value of $f=1.2 \times10^{-5}$ with a 6\% accuracy.  Unfortunately, this measurement disagrees by about a factor of 2.5 from essentially all other measurements and calculations.  We thus decided to renormalize the set to our calculated value for $n=9$. The resulting values have a striking agreement with both our calculations through $n=50$ and the other experimental measurements.  The agreement with Filippov \cite{Filippov1933} and, more significantly, with Mazing and Serapinas \cite{Mazing1969}, who covered the $n=18-30$ range, is also excellent.  It seems unfortunate that the high accuracy works of Huang and Wang \cite{Huang1981} and Sandner et al.\ \cite{Sandner1981} were published both in May 1981 and the authors seem to not have been aware of the one another's work, so that they could connect the results.

Our calculated optical oscillator strengths are presented in Tables \ref{table: K-4s-np_df_dE} and \ref{table: K-4s-np_rho}, and compared with available theoretical and experimental  values for up to $n=15$.  We notice an overall unanimity on the value of the resonant transition $4s-4p$, but a gradual deviation develops as $n$ increases.  These Tables include some data obtained from the absolute electrical dipole matrix elements published by Safronova et al.\ \cite{Safronova2013}.  Our values agree well with those of Weisheit \cite{Weisheit1972} and of Migdalek and Kim \cite{Migdalek1998}, but deviate from those of Safronova et al.\ \cite{Safronova2013} and Nandy et al.\ \cite{Nandy2012} for increasing $n$. We do not know the origin of this departure as we do not know how accurate the latter two treatments are for atomic states with large principal quantum numbers, which have their wavefunction maximum at distances beyond 100 atomic units. 

The theoretical results of Migdalek and Kim \cite{Migdalek1998} seem to agree well with ours over the entire common range of $n$, although a gradual departure is observed above about $n=8$; they also disagree with the values of Refs. \cite{Mazing1969, Shabanova1984, Huang1981}.  Of all calculations, the results using the Ref.\ \cite{Safronova2013} numbers seem to be the ones more ``out of synch'' with the others at increasing $n$; we cannot assess the origin of this disagreement; note that their inferred threshold value for $df/dE$ is much larger than all the theoretical and experimental results (shown in Table \ref{table: K-4s-cs_thr}).  
%(Note that the cross section values and oscillator strength distributions are connected as $\sigma { }(\text{Mb})=0.1240 df/dE{\ } [\text{Ry}]^{-1}$.)

\begin{table}[t]\small
\centering
\caption{\label{table: K-4s-np_df_dE}Comparison of calculated and observed oscillator strengths  for the K $4s$$\rightarrow$$\text{n}p_{3/2},\text{n}p_{1/2}$ transitions. CET, this work; MK, Ref.\ \cite{Migdalek1998}; SSC, Ref.\ \cite{Safronova2013}; NSSS, Ref.\ \cite{Nandy2012}; SK,  Ref.\ \cite{Shabanova1984}.}  %\\

\begin{tabular}{rc *{5}c}
\midrule\midrule
n&{\ \ } {n$^*$}&{\ \ } {CET}  &{\ \ }{MK}	     &{\ \ } {SCC}        &{\ \  }{NSSS}	&{\ \  }{SK}	 \\
\midrule
\multicolumn{7}{c}{$f(\text{n}s$$\rightarrow$$\text{n}p_{3/2}$)}\\
\midrule
4	&{\ }	2.235	&{\ }	6.55-1	&{\ }	6.69-1	&{\ }	6.68-1	&{\ }	6.72-1	&{\ }	6.52-1	\\
5	&{\ }	3.266	&{\ }	5.75-3	&{\ }	5.63-3	&{\ }	6.33-3	&{\ }	6.48-3	&{\ }	5.54-3	\\
6	&{\ }	4.276	&{\ }	6.27-4	&{\ }	5.75-4	&{\ }	7.88-4	&{\ }	7.63-4	&{\ }	5.76-4	\\
7	&{\ }	5.281	&{\ }	1.47-4	&{\ }	1.26-4	&{\ }	1.90-4	&{\ }	1.92-4	&{\ }	1.52-4	\\
8	&{\ }	6.283	&{\ }	5.09-5	&{\ }	4.05-5	&{\ }	7.42-5	&{\ }		&{\ }	4.56-5	\\
9	&{\ }	7.285	&{\ }	2.24-5	&{\ }	1.67-5	&{\ }	3.48-5	&{\ }		&{\ }	1.82-5	\\
10	&{\ }	8.286	&{\ }	1.16-5	&{\ }	8.14-6	&{\ }	2.02-5	&{\ }		&{\ }		\\
11	&{\ }	9.286	&{\ }	6.67-6	&{\ }	4.47-6	&{\ }		&{\ }		&{\ }		\\
12	&{\ }	10.287	&{\ }	4.19-6	&{\ }	2.68-6	&{\ }		&{\ }		&{\ }		\\
13	&{\ }	11.287	&{\ }	2.80-6	&{\ }	1.73-6	&{\ }		&{\ }		&{\ }		\\
14	&{\ }	12.288	&{\ }	1.97-6	&{\ }	1.17-6	&{\ }		&{\ }		&{\ }		\\
15	&{\ }	13.288	&{\ }	1.44-6	&{\ }	8.31-7	&{\ }		&{\ }		&{\ }		\\
\midrule
\multicolumn{7}{c}{$f(\text{n}s$$\rightarrow$$\text{n}p_{1/2}$)} \\
\midrule
4	&{\ }	2.232	&{\ }	3.26-1	&{\ }	3.33-1	&{\ }	3.33-1	&{\ }	3.34-1	&{\ }	3.24-1	\\
5	&{\ }	3.263	&{\ }	2.60-3	&{\ }	2.59-3	&{\ }	2.80-3	&{\ }	2.96-3	&{\ }	2.58-3	\\
6	&{\ }	4.273	&{\ }	2.59-4	&{\ }	2.45-4	&{\ }	3.08-4	&{\ }	3.31-4	&{\ }	2.53-4	\\
7	&{\ }	5.278	&{\ }	5.56-5	&{\ }	4.92-5	&{\ }	6.65-5	&{\ }	7.88-5	&{\ }	6.08-5	\\
8	&{\ }	6.280	&{\ }	1.77-5	&{\ }	1.46-5	&{\ }	2.21-5	&{\ }		&{\ }	1.72-5	\\
9	&{\ }	7.282	&{\ }	7.25-6	&{\ }	5.56-6	&{\ }	9.33-6	&{\ }		&{\ }	6.45-6	\\
10	&{\ }	8.283	&{\ }	3.50-6	&{\ }	2.51-6	&{\ }	4.99-6	&{\ }		&{\ }		\\
11	&{\ }	9.283	&{\ }	1.91-6	&{\ }	1.28-6	&{\ }		&{\ }		&{\ }		\\
12	&{\ }	10.284	&{\ }	1.14-6	&{\ }	7.21-7	&{\ }		&{\ }		&{\ }		\\
13	&{\ }	11.284	&{\ }	7.32-7	&{\ }	4.37-7	&{\ }		&{\ }		&{\ }		\\
14	&{\ }	12.285	&{\ }	4.96-7	&{\ }	2.82-7	&{\ }		&{\ }		&{\ }		\\
15	&{\ }	13.285	&{\ }	3.51-7	&{\ }	1.91-7	&{\ }		&{\ }		&{\ }		\\
\bottomrule
\end{tabular}
\end{table}

\begin{table}[h]\small
\centering%\label{Table2}
\caption{\label{table: K-4s-np_rho}Comparison of calculated and observed doublet sum and ratio of oscillator strengths  for the K $4s$$\rightarrow$$\text{n}p_{3/2},\text{n}p_{1/2}$ transitions. CET, this work; MK, Ref.\ \cite{Migdalek1998}; SSC, Ref.\ \cite{Safronova2013}; NSSS, Ref.\ \cite{Nandy2012}; W, Ref.\ \cite{Weisheit1972}; SK,  Ref.\ \cite{Shabanova1984}.}% \\

\begin{tabular}{rc *{6}c}
\midrule\midrule
n&{\ \ } {CET}  &{\ \ }{MK}	     &{\ \ } {SCC}        &{\ \  }{NSSS}	&{\ \  }{W}&{\ \  }{SK}&{\ \  }{HW}	 \\
\midrule
\multicolumn{8}{c}{$f(\text{n}p_{3/2})+f(\text{n}p_{1/2}$)}\\
\midrule
	
4	&{\ }	9.80-1	&{\ }	1.00+0	&{\ }	1.00+0	&{\ }	1.01+0	&{\ }	9.73-1	&{\ }	9.76-1&{\ }	 	\\
5	&{\ }	8.34-3	&{\ }	8.22-3	&{\ }	9.12-3	&{\ }	9.44-3	&{\ }	8.38-3	&{\ }	8.12-3&{\ } 	\\
6	&{\ }	8.86-4	&{\ }	8.20-4	&{\ }	1.10-3	&{\ }	1.09-3	&{\ }	8.23-4	&{\ }	8.29-4&{\ } 	\\
7	&{\ }	2.02-4	&{\ }	1.75-4	&{\ }	2.56-4	&{\ }	2.71-4	&{\ }	1.78-4	&{\ }	2.13-4&{\ }	1.85-4	\\
8	&{\ }	6.86-5	&{\ }	5.51-5	&{\ }	9.64-5	&{\ }		     &{\ }	5.74-5	&{\ }	6.28-5&{\ }	7.11-5	\\
9	&{\ }	2.96-5	&{\ }	2.23-5	&{\ }	4.41-5	&{\ }		     &{\ }	2.38-5	&{\ }	2.47-5&{\ }	2.96-5	\\
10	&{\ }	1.51-5	&{\ }	1.07-5	&{\ }	2.52-5	&{\ }		     &{\ }	1.17-5	&{\ }	1.11-5&{\ }	1.53-5	\\
11	&{\ }	8.59-6	&{\ }	5.75-6	&{\ }		     &{\ }		     &{\ }	6.46-6	&{\ }	5.87-6&{\ }	9.02-6	\\
12	&{\ }	5.33-6	&{\ }	3.40-6	&{\ }		     &{\ }		     &{\ }	3.92-6	&{\ }	3.73-6&{\ }	5.59-6	\\
13	&{\ }	3.53-6	&{\ }	2.17-6	&{\ }		     &{\ }		     &{\ }	2.55-6	&{\ }	2.64-6&{\ }	3.68-6	\\
14	&{\ }	2.46-6	&{\ }	1.45-6	&{\ }		     &{\ }		     &{\ }	1.72-6	&{\ }	1.90-6&{\ }	2.60-6	\\
15	&{\ }	1.79-6	&{\ }	1.02-6	&{\ }		     &{\ }		     &{\ }	1.25-6	&{\ }	         &{\ }1.88-6	\\
\midrule
\multicolumn{8}{c}{$\rho=f(\text{n}p_{3/2})/f(\text{n}p_{1/2}$)}\\
\midrule
4	&{\ }	2.008	&{\ }	2.009	&{\ }	2.015	&{\ }	2.012	&{\ }&{\ }	2.012&	\\
5	&{\ }	2.214	&{\ }	2.174	&{\ }	2.269	&{\ }	2.189	&{\ }&{\ }	2.147	\\
6	&{\ }	2.420	&{\ }	2.347	&{\ }	2.567	&{\ }	2.305	&{\ }&{\ }	2.277	\\
7	&{\ }	2.641	&{\ }	2.561	&{\ }	2.855	&{\ }	2.437	&{\ }&{\ }	2.500&{\ }	2.367	\\
8	&{\ }	2.867	&{\ }	2.774	&{\ }	3.360	&{\ }		     &{\ }&{\ }	2.651&{\ }	2.526	\\
9	&{\ }	3.088	&{\ }	3.004	&{\ }	3.730	&{\ }		     &{\ }&{\ }	2.822&{\ }	2.676	\\
10	&{\ }	3.298	&{\ }	3.243	&{\ }	4.066	&{\ }		     &{\ }&{\ }	       &{\ }	2.883   \\
11	&{\ }	3.492	&{\ }	3.492	&{\ }              &	                &{\ }&{\ }  	  &{\ }	3.134   \\
12	&{\ }	3.668	&{\ }	3.717	&{\ }              &	                &{\ }&{\ }  	  &{\ }	3.205   \\ 
13	&{\ }	3.826	&{\ }	3.959	&{\ }              &	                &{\ }&{\ }  	  &{\ }	3.398   \\
14	&{\ }	3.966	&{\ }	4.149	&{\ }              &	                &{\ }&{\ }  	  &{\ }	3.501   \\
15	&{\ }	4.091	&{\ }	4.351	&{\ }              &	                &{\ }&{\ }  	  &{\ }	3.592   \\
\bottomrule
\end{tabular}
\end{table}

\begin{table}[b]\small
\centering%\label{Table3}
\caption{\label{table: K-4s-cs_thr}Comparison of observed and calculated threshold cross section values, $\sigma_\text{th}$ (Mb), for the K($4s$) photoionization.}
\begin{tabular}{dcdc}
\midrule\midrule
\multicolumn{2}{c}{Experimental}&\multicolumn{2}{c}{Theoretical}\\
\midrule
{\ \ \ \ }0.012&{\ \ \ \ }\cite{Ditchburn1953}&{\ \ \ \ }0.006&{\ \ \ \ }\cite{Weisheit1972}\\
{\ \ \ \ }0.010(2)&{\ \ \ \ }\cite{Hudson1965}&{\ \ \ \ }0.011&{\ \ \ \ }\cite{Migdalek1998}\\
{\ \ \ \ }0.007(2)&{\ \ \ \ }\cite{Marr1968}&{\ \ \ \ }0.006&{\ \ \ \ }\cite{Petrov1999}\\
{\ \ \ \ }0.010(2)&{\ \ \ \ }\cite{Mazing1969}&{\ \ \ \ }0.01&{\ \ \ \ }\cite{Zatsarinny2010}\\
{\ \ \ \ }0.010(2)&{\ \ \ \ }\cite{Sandner1981}&{\ \ \ \ }0.00971&$\epsilon_{3/2+1/2}${\ \ }present\\
&&{\ \ \ \ }0.00815&$\epsilon_{3/2}${\ \ \ \ \ \ \ \ }present\\
&&{\ \ \ \ }0.00156&$\epsilon_{1/2}${\ \ \ \ \ \ \ \ }present\\
\bottomrule
\end{tabular}
\end{table}

%$$$$$$$$$$ fig.4 $$$$$$$$$$$$$$$$$$$$$$$$$$$$$
\begin{figure}[h]\centering\small\label{Fig4}
\includegraphics[width=7.6cm]{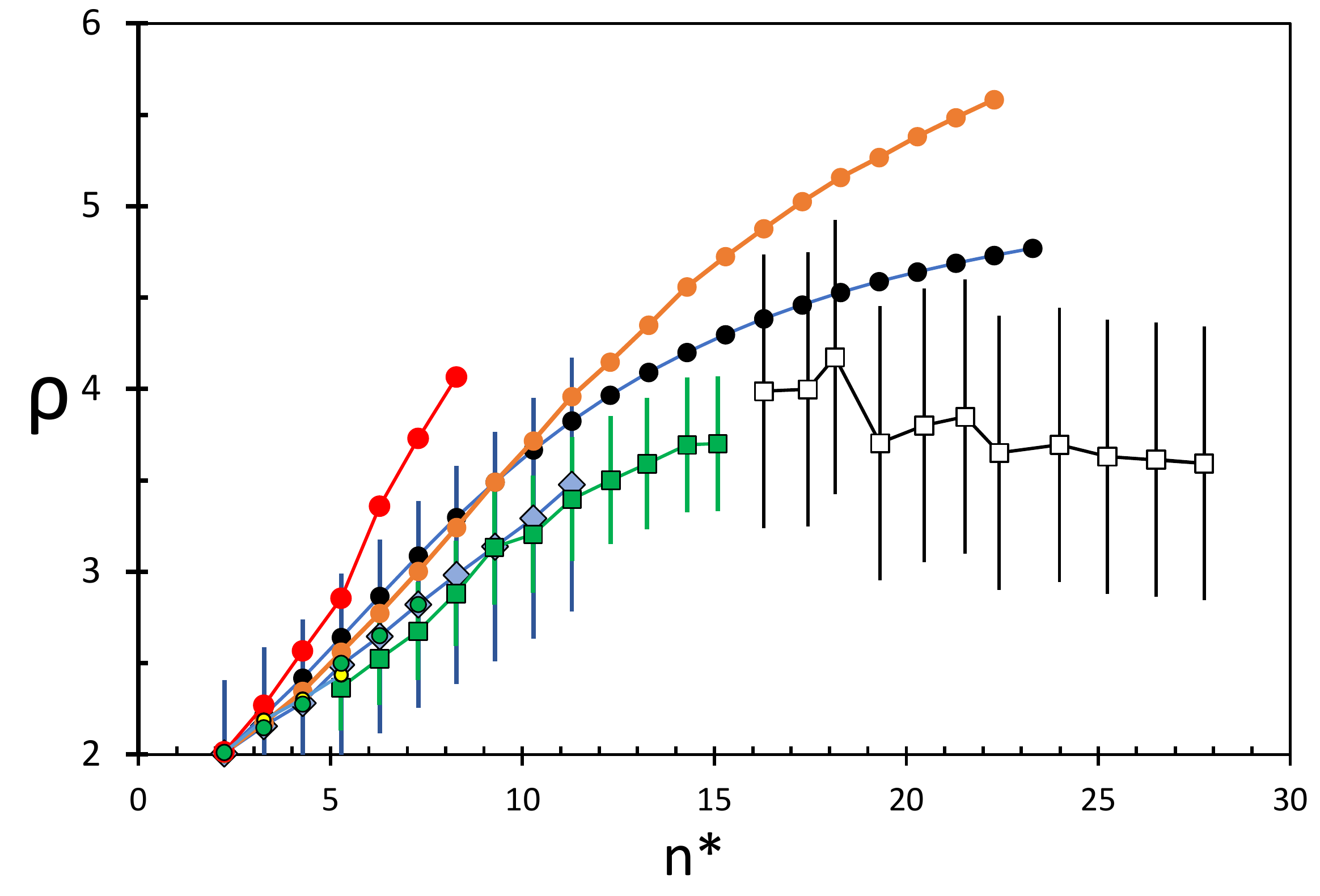}
\caption{\label{fig: K-4s-np_rho}   (Color online)  Ratio $\rho=f(np_{3/2})/f(np_{1/2})$. 
\textit{Experimental}:  $\mdblkdiamond$ (blue), Shabanova and Khlyustalov \cite{Shabanova1984}; $\mdblkcircle$ (green), data taken from NIST tabulation \cite{Wiese1969};   
$\mdblksquare$ (green), Huang and Wang \cite{Huang1981} for $n<17$; $\mdwhtsquare$, Huang and Wang \cite{Huang1981}, from deconvolution of the overlapping doublet components.
\textit{Theoretical}: $\mdblkcircle$  (red), using the matrix elements of Ref.\ \cite{Safronova2013}; $\mdblkcircle$  (yellow), Nandy et al.\ \cite{Nandy2012}; $\mdblkcircle$  (orange), Migdalek and Kim \cite{Migdalek1998}; $\mdblkcircle$ (black), present.}  
\end{figure}

Table \ref{table: K-4s-cs_thr} gives a detailed comparison of the various experimental and theoretical results for the threshold value of the photoionization cross section (and the inferred $df/dE$).  We see a coalescence of the results either around 0.006 Mb or 0.01 Mb.  Our data support the second value.

To assess the accuracy of the various measurements and calculations, we present in Fig.\ \ref{fig: K-4s-np_rho} the available results for the ratio $\rho=f(np_{3/2})/f(np_{1/2})$.  Agreement on this quantity can be the most differentiating between approaches.  We find our results to be very close to those of Migdalek and Kim \cite{Migdalek1998} to about $n=12$.  Agreement is also reasonable with the experimental results of Shabanova et al.\  \cite{Shabanova1984}. 
%; here we assume that since the individual oscillator strengths are of accuracy 20\%, their ratios are expected to have, at least, a similar error associated with them.  
Huang and Wang \cite{Huang1981} have provided the most extensive list of values, by direct observation of non-overlapping doublet members for $n<17$ and then by deconvolution of the doublet components for $n$ up to 30.  The agreement with our values through $n=20$ is very good. The experimental error increases after that and the obtained values of $\rho$ seem to change their monotonic $n-$dependent behavior.   The data from Ref.\ \cite{Safronova2013} seem, again, to be deviating the most from the other values, experimental, or calculated. The results of Nandy et  al.\ \cite{Nandy2012}, similar to the treatment of Ref.\ \cite{Safronova2013}, give $\rho$ values in the opposite direction from those of  Ref.\ \cite{Safronova2013}.  Our values appear to provide the best overall agreement with experiment.

In conclusion, we are confronted with the fact that all theories agree on the oscillator strength values at low principal quantum numbers $n$, but deviate from each other beyond about $n=8$.  Until a better resolution of this higher-$n$ behavior can be  reached, we can take advantage of the agreement of our calculations with the accurate experimental photoionization cross sections at threshold \cite{Sandner1981} and the excellent agreement with the measurements of Mazing and Serapinas\ \cite{Mazing1969} and Huang and Wang \cite{Huang1981} to establish a definitive self-consistent set of values for the entire principal series oscillator strength distribution. Table \ref{table: K-4s-np_f} presents the recommended values up to $n=50$, after which a linear fit of the $df/dE$ can yield the value for any $n$. The ionization limit ($n=\infty$) values are $\frac{df}{dE}(ns-np_{3/2})=1.01\times10^{-3}[\text{Ry}]^{-1}$ and $\frac{df}{dE}(ns-np_{1/2})=1.93\times10^{-4}[\text{Ry}]^{-1}$. 

\begin{acknowledgments}
I would like to dedicate this paper to the memory of Yong-Ki Kim, a kind and patient mentor. 
\end{acknowledgments}
\vfill

%\begin{table}
\begin{longtable}[h]{c|rrrrrr}
\caption{\label{table: K-4s-np_f} Oscillator strengths  for the K $4s$$\rightarrow$$\text{n}p_{3/2},\text{n}p_{1/2}$ transitions}\label{}  \\

\midrule \midrule
{$n$} &  $n^*${\ \ \ \ } &$np_{3/2}${\ \ \ \ } &  $n^*${\ \ \ \ } &$\text{n}p_{1/2}${\ \ \ \ }  &$\text{n}p_{3/2+1/2}$ {\  }& $\rho${\ \ \ }\\
\midrule
\endfirsthead
4	&{\ \ }	2.235	&{\ \ }	6.545-1	&{\ \ }	2.232	&{\ \ }	3.260-1	&{\ \ }	9.805-1	&{\ \ }	2.01	\\
5	&{\ \ }	3.266	&{\ \ }	5.748-3	&{\ \ }	3.263	&{\ \ }	2.596-3	&{\ \ }	8.344-3	&{\ \ }	2.21	\\
6	&{\ \ }	4.276	&{\ \ }	6.268-4	&{\ \ }	4.273	&{\ \ }	2.590-4	&{\ \ }	8.858-4	&{\ \ }	2.42	\\
7	&{\ \ }	5.281	&{\ \ }	1.468-4	&{\ \ }	5.278	&{\ \ }	5.560-5	&{\ \ }	2.024-4	&{\ \ }	2.64	\\
8	&{\ \ }	6.283	&{\ \ }	5.086-5	&{\ \ }	6.280	&{\ \ }	1.774-5	&{\ \ }	6.861-5	&{\ \ }	2.87	\\
9	&{\ \ }	7.285	&{\ \ }	2.238-5	&{\ \ }	7.282	&{\ \ }	7.246-6	&{\ \ }	2.962-5	&{\ \ }	3.09	\\
10	&{\ \ }	8.286	&{\ \ }	1.155-5	&{\ \ }	8.283	&{\ \ }	3.502-6	&{\ \ }	1.505-5	&{\ \ }	3.30	\\
11	&{\ \ }	9.286	&{\ \ }	6.674-6	&{\ \ }	9.283	&{\ \ }	1.911-6	&{\ \ }	8.585-6	&{\ \ }	3.49	\\
12	&{\ \ }	10.287	&{\ \ }	4.190-6	&{\ \ }	10.284	&{\ \ }	1.143-6	&{\ \ }	5.333-6	&{\ \ }	3.67	\\
13	&{\ \ }	11.287	&{\ \ }	2.802-6	&{\ \ }	11.284	&{\ \ }	7.324-7	&{\ \ }	3.534-6	&{\ \ }	3.83	\\
14	&{\ \ }	12.288	&{\ \ }	1.968-6	&{\ \ }	12.285	&{\ \ }	4.961-7	&{\ \ }	2.464-6	&{\ \ }	3.97	\\
15	&{\ \ }	13.288	&{\ \ }	1.437-6	&{\ \ }	13.285	&{\ \ }	3.512-7	&{\ \ }	1.788-6	&{\ \ }	4.09	\\
16	&{\ \ }	14.288	&{\ \ }	1.083-6	&{\ \ }	14.285	&{\ \ }	2.578-7	&{\ \ }	1.341-6	&{\ \ }	4.20	\\
17	&{\ \ }	15.288	&{\ \ }	8.377-7	&{\ \ }	15.285	&{\ \ }	1.949-7	&{\ \ }	1.033-6	&{\ \ }	4.30	\\
18	&{\ \ }	16.288	&{\ \ }	6.622-7	&{\ \ }	16.285	&{\ \ }	1.510-7	&{\ \ }	8.132-7	&{\ \ }	4.38	\\
19	&{\ \ }	17.288	&{\ \ }	5.332-7	&{\ \ }	17.285	&{\ \ }	1.195-7	&{\ \ }	6.527-7	&{\ \ }	4.46	\\
20	&{\ \ }	18.288	&{\ \ }	4.361-7	&{\ \ }	18.285	&{\ \ }	9.631-8	&{\ \ }	5.325-7	&{\ \ }	4.53	\\
21	&{\ \ }	19.289	&{\ \ }	3.616-7	&{\ \ }	19.285	&{\ \ }	7.882-8	&{\ \ }	4.405-7	&{\ \ }	4.59	\\
22	&{\ \ }	20.289	&{\ \ }	3.034-7	&{\ \ }	20.286	&{\ \ }	6.538-8	&{\ \ }	3.688-7	&{\ \ }	4.64	\\
23	&{\ \ }	21.289	&{\ \ }	2.573-7	&{\ \ }	21.286	&{\ \ }	5.487-8	&{\ \ }	3.122-7	&{\ \ }	4.69	\\
24	&{\ \ }	22.289	&{\ \ }	2.202-7	&{\ \ }	22.286	&{\ \ }	4.653-8	&{\ \ }	2.667-7	&{\ \ }	4.73	\\
25	&{\ \ }	23.289	&{\ \ }	1.900-7	&{\ \ }	23.286	&{\ \ }	3.983-8	&{\ \ }	2.298-7	&{\ \ }	4.77	\\
26	&{\ \ }	24.289	&{\ \ }	1.651-7	&{\ \ }	24.286	&{\ \ }	3.437-8	&{\ \ }	1.995-7	&{\ \ }	4.80	\\
27	&{\ \ }	25.289	&{\ \ }	1.445-7	&{\ \ }	25.286	&{\ \ }	2.989-8	&{\ \ }	1.744-7	&{\ \ }	4.83	\\
28	&{\ \ }	26.289	&{\ \ }	1.272-7	&{\ \ }	26.286	&{\ \ }	2.616-8	&{\ \ }	1.534-7	&{\ \ }	4.86	\\
29	&{\ \ }	27.289	&{\ \ }	1.126-7	&{\ \ }	27.286	&{\ \ }	2.304-8	&{\ \ }	1.357-7	&{\ \ }	4.89	\\
30	&{\ \ }	28.289	&{\ \ }	1.002-7	&{\ \ }	28.286	&{\ \ }	2.040-8	&{\ \ }	1.206-7	&{\ \ }	4.91	\\
31	&{\ \ }	29.289	&{\ \ }	8.957-8	&{\ \ }	29.286	&{\ \ }	1.816-8	&{\ \ }	1.077-7	&{\ \ }	4.93	\\
32	&{\ \ }	30.289	&{\ \ }	8.041-8	&{\ \ }	30.286	&{\ \ }	1.624-8	&{\ \ }	9.665-8	&{\ \ }	4.95	\\
33	&{\ \ }	31.289	&{\ \ }	7.247-8	&{\ \ }	31.286	&{\ \ }	1.458-8	&{\ \ }	8.705-8	&{\ \ }	4.97	\\
34	&{\ \ }	32.289	&{\ \ }	6.555-8	&{\ \ }	32.286	&{\ \ }	1.315-8	&{\ \ }	7.870-8	&{\ \ }	4.98	\\
35	&{\ \ }	33.289	&{\ \ }	5.949-8	&{\ \ }	33.286	&{\ \ }	1.190-8	&{\ \ }	7.139-8	&{\ \ }	5.00	\\
36	&{\ \ }	34.289	&{\ \ }	5.417-8	&{\ \ }	34.286	&{\ \ }	1.081-8	&{\ \ }	6.498-8	&{\ \ }	5.01	\\
37	&{\ \ }	35.289	&{\ \ }	4.947-8	&{\ \ }	35.286	&{\ \ }	9.844-9	&{\ \ }	5.931-8	&{\ \ }	5.03	\\
38	&{\ \ }	36.289	&{\ \ }	4.530-8	&{\ \ }	36.286	&{\ \ }	8.995-9	&{\ \ }	5.430-8	&{\ \ }	5.04	\\
39	&{\ \ }	37.289	&{\ \ }	4.160-8	&{\ \ }	37.286	&{\ \ }	8.241-9	&{\ \ }	4.984-8	&{\ \ }	5.05	\\
40	&{\ \ }	38.289	&{\ \ }	3.828-8	&{\ \ }	38.286	&{\ \ }	7.570-9	&{\ \ }	4.586-8	&{\ \ }	5.06	\\
41	&{\ \ }	39.289	&{\ \ }	3.532-8	&{\ \ }	39.286	&{\ \ }	6.971-9	&{\ \ }	4.229-8	&{\ \ }	5.07	\\
42	&{\ \ }	40.289	&{\ \ }	3.266-8	&{\ \ }	40.286	&{\ \ }	6.435-9	&{\ \ }	3.909-8	&{\ \ }	5.07	\\
43	&{\ \ }	41.289	&{\ \ }	3.025-8	&{\ \ }	41.286	&{\ \ }	5.952-9	&{\ \ }	3.621-8	&{\ \ }	5.08	\\
44	&{\ \ }	42.289	&{\ \ }	2.809-8	&{\ \ }	42.286	&{\ \ }	5.518-9	&{\ \ }	3.360-8	&{\ \ }	5.09	\\
45	&{\ \ }	43.289	&{\ \ }	2.612-8	&{\ \ }	43.286	&{\ \ }	5.125-9	&{\ \ }	3.124-8	&{\ \ }	5.10	\\
46	&{\ \ }	44.289	&{\ \ }	2.434-8	&{\ \ }	44.286	&{\ \ }	4.768-9	&{\ \ }	2.910-8	&{\ \ }	5.10	\\
47	&{\ \ }	45.289	&{\ \ }	2.271-8	&{\ \ }	45.286	&{\ \ }	4.445-9	&{\ \ }	2.716-8	&{\ \ }	5.11	\\
48	&{\ \ }	46.289	&{\ \ }	2.123-8	&{\ \ }	46.286	&{\ \ }	4.150-9	&{\ \ }	2.538-8	&{\ \ }	5.12	\\
49	&{\ \ }	47.289	&{\ \ }	1.987-8	&{\ \ }	47.286	&{\ \ }	3.881-9	&{\ \ }	2.375-8	&{\ \ }	5.12	\\
50	&{\ \ }	48.289	&{\ \ }	1.863-8	&{\ \ }	48.286	&{\ \ }	3.635-9	&{\ \ }	2.227-8	&{\ \ }	5.13	\\
$\infty$&&&&&&{\ \ }5.24\\
\bottomrule
\end{longtable}

\end{document}